\documentclass[letterpaper,11pt]{article}
\usepackage{geometry} 
\usepackage[utf8]{inputenc}
\usepackage[T1]{fontenc}
\usepackage{lmodern}

\usepackage[dvipsnames,table]{xcolor}
\usepackage{graphicx}
\graphicspath{{./}{figures/}}
\usepackage{hyperref} 
\hypersetup{colorlinks = true,
            linkcolor=black,
            urlcolor=RoyalBlue,
            citecolor=black,
            }
\usepackage{url}
\usepackage{gensymb}
\usepackage{fancyhdr}
\usepackage{wrapfig}
\usepackage{sidecap}
\usepackage{natbib}  
\usepackage{amsmath}
\usepackage{amssymb}
\usepackage[nameinlink,noabbrev]{cleveref}
\usepackage{tcolorbox}
\usepackage{longtable}
\usepackage{enumitem}
\usepackage{floatrow}
\newfloatcommand{capbtabbox}{table}[][\FBwidth]
\usepackage{lipsum}
\usepackage[displaymath, mathlines]{lineno}
\modulolinenumbers[2]
\usepackage{float}
\usepackage{pdfpages} 
\usepackage{todonotes}
\usepackage{sectsty}
\usepackage{authblk}
\usepackage{xspace}

\sectionfont{\fontsize{12}{15}\selectfont}
\subsectionfont{\fontsize{12}{15}\selectfont}

\definecolor{lightgray}{gray}{0.9}
\date{\vspace{-5ex}}

\begin{document}

\noindent
\textbf{\large \textit{Roman} CCS White Paper}

\vspace{2em}
\begin{center}
{\Large Measuring Type Ia Supernovae Discovered in the \textit{Roman} High Latitude Time Domain Survey}
\end{center}

\vspace{1.5em}
\noindent
\textbf{\textit{Roman} Core Community Survey:} High Latitude Time Domain Survey

\vspace{0.5em}
\noindent
\textbf{Scientific Categories:} Stellar Physics and Stellar Types

\vspace{0.5em}
\noindent
\textbf{Additional scientific keywords:} Supernovae, Cosmology, Dark energy, Infrared Photometry

\vspace{1em}
\noindent
\textbf{Submitting Author:}\\
Rebekah Hounsell, University of Maryland Baltimore County/ NASA Goddard Space Flight Center (rebekah.a.hounselle@nasa.gov)

\vspace{0.5em}
\noindent
\textbf{List of contributing authors:}\\
Dan Scolnic, Duke University, (daniel.scolnic@duke.edu) \\
Dillon Brout, Boston University\\
Benjamin Rose, Baylor University (Ben\_Rose@baylor.edu)\\
Ori Fox, Space Telescope Science Institute (ofox@stsci.edu)\\
Masao Sako, University of Pennsylvania (masao@sas.upenn.edu)\\
Phillip Macias, UC Santa Cruz (pmacias@ucsc.edu)\\
Bhavin Joshi, Johns Hopkins University (bjoshi5@jhu.edu)\\
Susana Desutua, NIST (susana.deustua@nist.gov)\\
David Rubin, UH (drubin@hawaii.edu)\\
Stefano Casertano, STScI (stefano@stsci.edu)\\
Saul Perlmutter, University of California, Berkeley (saul@lbl.gov)\\
Greg Aldering, Lawrence Berkeley National Lab (galdering@lbl.gov)\\
Kaisey Mandel, University of Cambridge (kmandel@ast.cam.ac.uk)\\
Megan Sosey, STScI (sosey@stsci.edu)\\
Nao Suzuki, Lawrence Berkeley National Lab (nsuzuki@lbl.gov)\\
Russell Ryan, STScI (rryan@stsci.edu) \\

\vspace{1em}
\noindent
\textbf{Abstract:} 

We motivate the cosmological science case of measuring Type Ia supernovae with the \textit{Nancy Grace Roman Space Telescope} as part of the High Latitude Time Domain Survey.  We discuss previously stated requirements for the science, and a baseline survey strategy.  We discuss the various areas that must still be optimized and point to the other white papers that consider these topics in detail.  Overall, the baseline case should enable an exquisite measurement of dark energy using SNe~Ia from $z=0.1$ to $z>2$, and further optimization should only strengthen this once-in-a-generation experiment.

\thispagestyle{empty}
\newpage
\setcounter{page}{1}


\section{Introduction}
The \textit{Nancy Grace Roman Space Telescope} (\textit{Roman}) is NASA’s next large flagship mission, due for launch in late 2026. One of the mission’s key objectives is to determine the expansion history of the universe and to test possible explanations for its apparent acceleration, including dark energy and modifications to general relativity. The three main cosmological probes it will utilize are Weak Lensing, Galaxy Redshift measurements and Type Ia supernovae (SN~Ia). In order to discover and measure SNe~Ia, the mission will conduct a generation-defining experiment in time-domain astronomy via a Core Community survey called the High Latitude Time Domain Survey (HLTDS). 

\subsection{Constraining cosmological parameters with Roman Measurements of SNe Ia}

The measurement of SNe~Ia distances across a wide range of redshift is a powerful and complementary cosmological probe compared to the weak lensing and galaxy redshift probes.  

A SN~Ia survey has been planned within the \textit{Roman} mission since its inception, and there are clear requirements for its implementation.  One of the major requirements is to obtain a large sample of SN~Ia: $\geq$100 per $\Delta$z=0.1 bin over 0.2 $\leq$ z $\leq$ 1.7. 
For success, Near Infra-Red (NIR) images must be obtained in multiple bands, and with a $\sim$5-day cadence to ensure adequate SN light curve coverage. 

The depth of the observations, size of the fields, and location  have been left for optimization studies and will be the subject of future work within the community.

An additional requirement is on the Dark Energy Task Force Figure of Merit \citep[DETF-FoM,][]{Albrecht06} from the combination of the cosmological constraints.  The stated requirement is that \textit{Roman} must be a Stage IV Dark Energy mission. While it is difficult to predict the complementary/orthogonality of the SN~Ia constraints with the other probes, a simple metric is that constraints for each probe must be $2-3\times$ better than the Stage III experiments (e.g. the Dark Energy Supernova program).

\subsection{What will make the SN~Ia survey successful}

The constraining power from SN~Ia depends on a differential measurement of the brightness of SNe~Ia across a redshift range. Improving constraints on cosmological parameters can be enabled by increasing the redshift range, increasing the number of SNe~Ia observed, increasing the precision of each SN~Ia measurement, and limiting the impact of systematic uncertainties.

The various parts of the constraining power play off each other: there are limited returns when the statistical precision of all the SNe~Ia reaches below a systematic floor and there are limited returns when the precision of a single distance measurement is significantly below the intrinsic scatter of a SN~Ia brightness. Given these considerations, the community has proposed collecting a robust and statistically
significant / complete sample of cosmology-worthy SN~Ia. The proposed depth is such that
within each redshift bin, the accuracy is still limited by the intrinsic scatter in individual SNe (roughly
0.1 mag and 0.06 mag for imaging and spectroscopy, respectively); the systematic
accuracy of the flux scale is required to be better than 3 mmag. 

There are two main ways of constraining distances. The first is from measurements of the light-curve photometry and has been used in recent cosmological analyses \citep[e.g., ][]{Scolnic22, DES3}. A second approach, developed by the SNfactory \citep{SNFactory} is to compare SNe~Ia spectra directly, as in \citet{BooneTE2021ApJ...912...71B, Stein_2022ApJ...935....5S}. This alternate approach is enabled by prism measurements, and HLTDS which have an increased fraction of prism time are discussed in \cite{Rose21} and the white paper by Aldering et. al., entitled \textit{"Balanced Prism Plus Filter Cadence in
the High Latitude Time Domain Survey Core Community Survey"} (hereafter referred to as the Aldering prism paper).

An additional consideration is the acquisition of redshifts. Almost all cosmological analyses with SNe~Ia have relied on spectroscopic redshifts, either from the host galaxy or the SN itself. These measurements will be enabled by the \textit{Roman} prism, grism, or external sources e.g., ground-based follow-up. In addition to spectroscopic redshifts, the usage of photometric redshifts \citep[e.g.,][]{Photoz} is viable in certain situations. The use of photometric redshifts for SN~Ia cosmology is still in development, with continued investigation of systematics control and contamination \citep{Chen2022}, but the potential redshifts acquired from photometry alone should be considered as part of the motivation of the observing strategy for the \textit{Roman} deep fields. This affects both field location, field depth, and number of filters used. 

\section{Survey Design}
In \citet{Rose21} an initial High Latitude Time Domain reference survey is presented; this survey design will be the focus of our white paper. While the 25\% prism, 75\% imaging HLTDS presented in \citet{Rose21} focuses on the acquisition of SN~Ia, it will also be extremely beneficial for all kinds of transient studies. This design utilizes six of the Wide Field Instrument (WFI) imaging filters (F062,F087,F106,F129,F158,F184 corresponding to R,Z,Y,J,H,F in Table~\ref{tab:25}) and the low resolution prism (P127).

\begin{table}[b]
\centering
\footnotesize
\begin{tabular}{|l l c|c c r r r|r|} 
\hline
Mode & Tier & z$_{targ}$ & Filters & Exp.Time+Overhead & No. of & Area & Time/Visit & Total\\
&  & &  & (s) & Pointings & (deg$^2$) & (hours)  & SNe~IA\\ 

\hline\hline
Imaging & Wide & 1.0 & R;Z;Y;J & 160;100;100;100 + 70x4 & 68 & 19.04 & 14.0 &8804 \\ 
Imaging & Deep & 1.7 & Y;J;H;F & 300;300;300;900 + 70x4 & 15 &  4.20 & 8.5 &3520  \\ 
\hline
\textbf{Subtotal Imaging} & & &  &  &  &  & \textbf{22.5} & \textbf{12324} \\ 
\hline
Spec & Wide & 1.0 & prism & \hfill 900 + 70~~~~ & 12 & 3.36 & 3.2 & 831  \\ 
Spec & Deep &  1.5 & prism & \hfill 3600 + 70~~~~ &  4 & 1.12 & 4.1 &652 \\ 
\hline
\textbf{Subtotal Spec} & &  &  &  &  &  & \textbf{7.3} &  \textbf{1483}  \\ 
\hline 
\end{tabular}
\caption{The High-latitude Time Domain initial reference survey taken from Table 1 of \citet{Rose21}. Note that $z_{targ}$ denotes the redshift where the average SNe~Ia at peak is observed with S/N=10 per exposure for imaging, and S/N=25 for spectroscopy}
\label{tab:25}
\end{table}

The six filters selected provide the necessary wavelength range (0.62--1.84~$\mu$m) required to capture SN~Ia light curves at peak across a broad range of redshift. Note that in all current investigations, the F213-band (K-band) has been excluded. While this band-pass may provide useful information, the thermal noise (4.52 counts per pixel\footnotemark) of this filter makes it expensive to use. Note that the K-band could be implemented successfully within an external survey or study as described in the white papers submitted by Gomez et. al., in \textit{"Characterizing Superluminous Supernovae with Roman"} and Fox et. al., in \textit{"An Extended Time-Domain Survey (eTDS) to Detect High-z Transients, Trace the First Stars, and Probe the Epoch of Reionization"}.

Multiple tiers are necessary such that specific redshift SN~Ia can be targeted by the correct filters, and exposure times tailored to specific S/N requirements. In this case, the wide tier targets SNe~Ia at z $\leq$ 1, and the deep SNe~Ia at z $\sim$ 1.7. Each exposure time is tailored to obtaining a SNe~Ia S/N of 10 at peak in the imaging filters and for the prism an integrated S/N of 25 in the V-band at rest frame when spectra +/- 5 days from peak are co-added. Such a design is supported by work conducted in \citet{Hounsell}, in which several two-tier survey strategies were examined and DETF-FoM values ranging from 133 to 352 (using optimistic uncertainties) were obtained.

As stated previously the initial reference survey uses the low resolution prism (0.75--1.80 $\mu$m) only 25$\%$ of the time. This value can of course be adjusted, and several alternatives are already discussed within ~\citet{Rose21}. The prism would be used for spectroscopic discovery and follow-up of SN~Ia. The impact of \textit{Roman's} prism on transient science is discussed in the Aldering prism paper, with additional discussion in the white paper by Rose et. al., entitled \textit{"Options to Increase the Coverage Area of Prism Time Series in the High-Latitude Time Domain Core Community Survey"}.

\footnotetext{\url{https://roman.gsfc.nasa.gov/science/WFI_technical.html}}

\subsection{Survey Duration and Cadence}
The initial reference surveys in \citet{Rose21} used the configuration as presented in \citet{Spergel15} and \citet{Hounsell} i.e., 30 hr visits, every five days, over two years, for a total observing time of six months. 

Six months is the minimal duration necessary to obtain a quality sample of SNe over a broad range of redshifts. Note however, that the typical light curve length at rest frame for a SN~Ia is $\sim$45~days.  At a redshift of $z=2$, this would be approaching 5~months of survey length. A longer duration survey would allow the procurement of further template images and spectra for subtraction from SNe outburst data and so further refinement of the sample. Less time would significantly impact the number of SN~Ia obtained per redshift bin and as such compromise one of the main mission objectives.

A cadence of $\sim$5 days is suitable for SN~Ia observed at both $z < 1$ and above. However, alternative cadences of 4/5 days for the wide field and 8/10 days for the deep have been investigated with promising results. Shorter cadences in the wide tier however would not be beneficial for the overall SN~Ia science goals as a separate part of the survey would have to be weakened and the benefit from faster sampling of light curves that are not particularly fast is limited. While potentially beneficial for other faster transients, a cadence faster than $\sim$5 days is not required to fully capture the SN~Ia evolution and to create a robust statistical sample of low z SNe. 
A cadence much longer than 8--10 days for the deep field could result in too sparse data points for mid-range redshift SNe. For further discussion about the cadence please see Rubin et. al., \textit{"Optimizing the Cadence at Fixed Depth"}.

\subsection{Field location}

The white paper presented by Rose et. al., entitled \textit{"Considerations for Selecting Fields for the Roman High-latitude Time Domain Core Community Survey"} presents a full discussion on the complexities of selecting field locations for the SN~Ia survey. The Aldering prism paper also looks at prism survey speed.

Selecting an appropriate field is a complex decision and reliant upon many factors such as Milky Way extinction, Zodiacal background, Continuous Viewing Zones, synergies with other mission/project surveys etc. Fields with high absolute ecliptic and galactic latitudes would therefore be preferable, and as such multiple fields per tier may be required such that an area is covered in both the northern and the southern hemispheres allowing for additional survey/ground based follow up. Having fields in different hemispheres would also allow for a jackknife re-sampling test. The location of potential \textit{Roman} SN~Ia fields are presented in Figure 1. In the southern hemisphere the possible field includes the Akari Deep Field South (ADFS)/Euclid Deep Field South (EDFS) region and in the northern, Extended Groth Strip (EGS), Supernova/Acceleration Probe North (SNAP-N), and European Large Area Infrared Space Observatory Survey-North 1 (ELAIS N-1) are good possibilities. 

\begin{figure}
    \centering
    \label{fig:field}
    \includegraphics[width=0.85\textwidth]{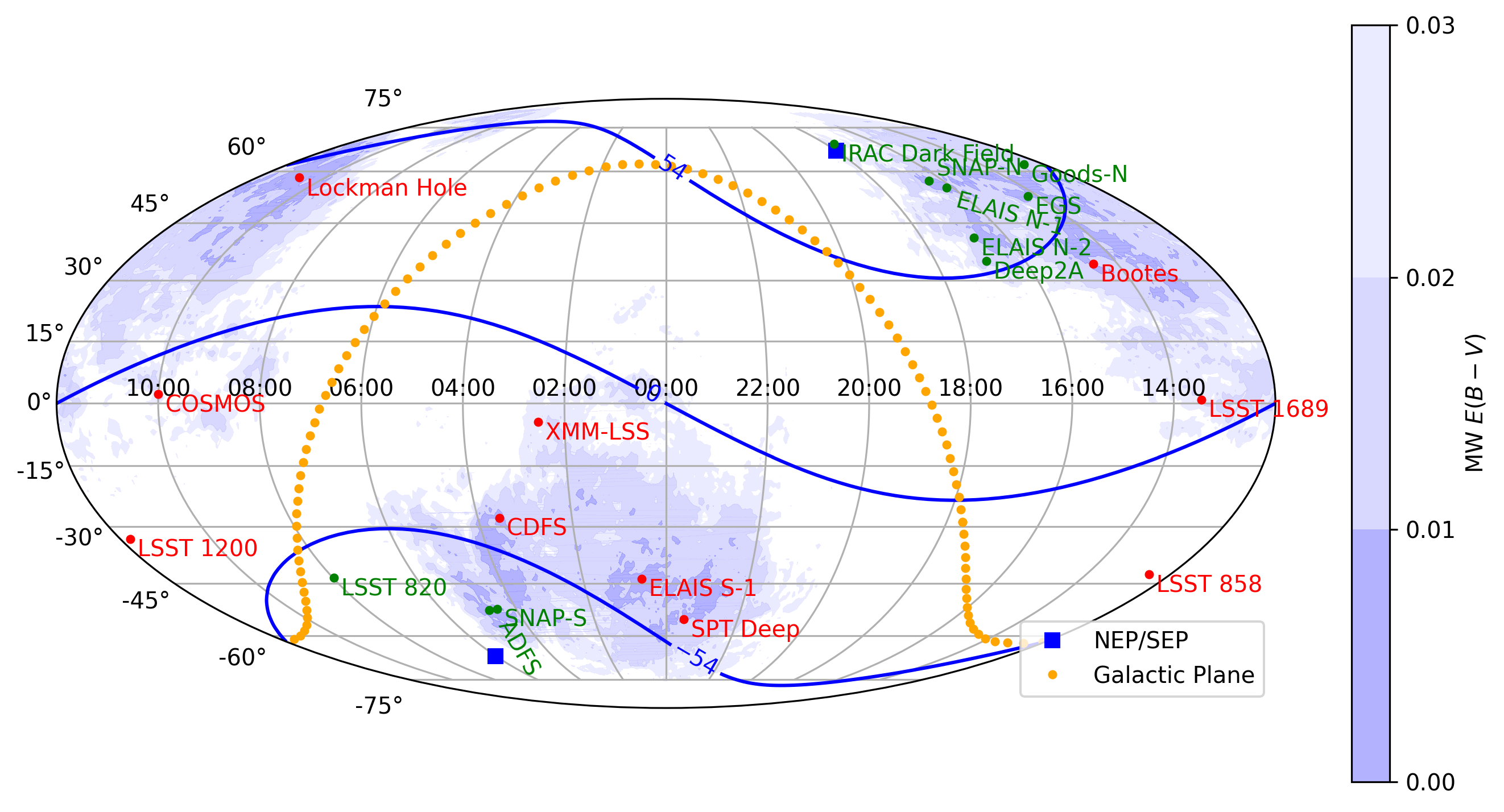}
    \caption{An all-sky plot of possible Roman time-domain fields. Green labels are in the current Roman CVZ ($\pm54^{\degree}$ off the ecliptic) and red labels are outside the CVZ. Low Milky Way dust extinction is shown as blue shading. Overall, the top field choices include Extended Groth Strip (EGS),Supernova/Acceleration Probe North (SNAP-N), European Large Area Infrared Space Observatory Survey-North 1 (ELAIS N-1), and Akari Deep Field South (ADFS)/Euclid Deep Field South (EDFS). Image adapted by B. Rose from Figure 1 in \citet{Rose21}.}
\end{figure}

\section{Conclusion}

From previous studies, there is a good baseline strategy for discovering and measuring SNe~Ia with Roman. There are still a number of open questions, like field location and filter allocation, and the community should figure out which questions would be beneficial early for advanced survey planning, and which questions should continue to be optimized due to better understanding of the instrument and further insights from ongoing studies from other surveys and analyses. 

\bibliographystyle{apj}
\bibliography{lib}

\end{document}